\documentclass[a4paper]{jpconf}
\pdfoutput=1
\usepackage{amsmath}
\usepackage{graphicx}
\usepackage{braket}

\begin{document}

\title{Note on a less ad hoc derivation of the second quantization formalism}

\author{Ning Wu}
\ead{wunwyz@gmail.com}
\address{Center for Quantum Technology Research, School of Physics, Beijing Institute of Technology, Beijing 100081, China}

\vspace{10pt}

\begin{abstract}
Second quantization is an essential topic in senior undergraduate and postgraduate level Quantum Mechanics course. However, it seems that there is a lack of transparent and natural derivation of this formalism from the first-quantization one in most existing textbooks. Without introducing the concept of a simple harmonic oscillator and taking the case of an assembly of identical fermions as an example, we provide a less ad hoc derivation of the second quantization formalism based on the equivalence between the fully antisymmetric many-fermion states and the Fock states.
\end{abstract}

%
%
%
%
%

\section{Introduction}
Second quantization is a useful and important tool for treating systems of many identical interacting particles. It is also an essential topic for upper-division and postgraduate physics students when they study Advanced Quantum Mechanics. According to the author's teaching experience, it is usually difficult for the students to fully understand the construction of the second quantization formalism from the wave-function description of the many-particle system, due to the reasons will be detailed below. The aim of this note is to provide a \emph{less ad hoc} derivation of the second quantization formalism based merely on the equivalence of the fully symmetric or antisymmetric many-particle states and the corresponding Fock states in the occupation number representation.
\par It is well known that the free classical electromagnetic field is described by an infinite set of uncoupled harmonic oscillators, which makes the second quantization of a multiple boson system natural by directly extending the formalism of a single harmonic oscillator to the multimode case. This is best illustrated in Dirac's classical book~\cite{Dirac} (see $\S$60 there) and is perhaps one reason why most textbooks on quantum mechanics often introduce the commutation or anticommutation relations among the particle creation and annihilation operators immediately after introducing the concept of a Fock space, see, for example, Refs.~\cite{Sakurai,Gottfried,Ballentine,Baym}. We argue that the knowledge of the algebraic solution of a harmonic oscillator is actually not necessary for the construction. On the contrary, it may even lead to confusions for the beginners on this topic. For instance, the students may get confused when considering $N$ identical fermions trapped in a harmonic oscillator potential: on the one hand, the creation and annihilation operators for the energy quanta of a single oscillator satisfy the (single-mode) bosonic commutation relations; on the other hand, the creation and annihilation operators for the real fermions in second quantization have to obey the fermionic anticommutation relations. The only useful idea we borrow from the harmonic oscillator is the concept of ``\emph{filling the vacuum}", which is also the basis for a Fock space.
\par As pointed out in the book by Sakurai \emph{et. al.}~\cite{Sakurai}, imposing these canonical commutation relations is actually ``\emph{very ad hoc}" and ``it is not possible to do a fully self-consistent treatment minimizing ad hoc assumptions without developing relativistic quantum field theory". In this note, we will provide a ``less ad hoc" derivation of the second quantization. Our development is based on:\\
\\
~ i) The symmetrization postulate,\\
~ ii) The concept of ``filling the vacuum" in the Fock space,\\
~ iii) The equivalence between the symmetrized or antisymmetrized many-particle wave function and the Fock state. \\
\\
We will focus on the case of Fermi statistics due to the simplicity in the normalization of the many-particle wave function.
\section{Equivalence between fully antisymmetric states and Fock states}\label{SecII}
\subsection{\textbf{Setup in the first-quantization language}}
\par Consider an interacting system composed of $N$ identical fermions. In the ``first quantization" language, we are forced to label the single-particle states for particle $i$ as $|\phi_\alpha\rangle_i$, where $i=1,2,\cdots,N$ and $\alpha$ is the index for the single particle states (SPSs). We assume there are totally $M$ SPSs and $M$ can be either a finite (e.g., the two spin states for a spin-1/2 particle) or an infinite number (e.g., the eigenstates of a simple harmonic oscillator), depending on the physical system being considered.
\par Assuming that the $M$ SPSs $\{|\phi_\alpha\rangle_i|\alpha=1,2,\cdots,M\}$ form a complete set for particle $i$, i.e,
\begin{eqnarray}
\sum^M_{\alpha=1}|\phi_\alpha\rangle_i~_i\langle\phi_\alpha|=1_i, 
\end{eqnarray} 
one complete set of the $N$-fermion system is simply the set of all product states,
\begin{eqnarray}
|\phi_{P_1}\rangle_1\otimes|\phi_{P_2}\rangle_2\otimes\cdots\otimes|\phi_{P_N}\rangle_N\equiv |\phi_{P_1}\rangle_1 |\phi_{P_2}\rangle_2 \cdots |\phi_{P_N}\rangle_N,\nonumber
\end{eqnarray}
where $P_i\in\{1,2,\cdots,M\}$ is the index of the SPS occupied by particle $i$. Since the $N$ fermions are identical, we can always choose $1\leq P_1\leq P_2\leq\cdots\leq P_N\leq M$. Below we shall call this particular product state $|\phi_{P_1}\rangle_1 |\phi_{P_2}\rangle_2 \cdots |\phi_{P_N}\rangle_N$, in which the SPS indices are ordered as $1\leq P_1\leq P_2\leq\cdots\leq P_N\leq M$, a \emph{reference state}.
\par The first-quantization Hamiltonian for an $N$-fermion system can generally be written as
\begin{eqnarray}
H&=&F^{(1)}+F^{(2)},\nonumber\\
F^{(1)}&=&\sum^N_{i=1}f^{(1)}_i,~F^{(2)}=\frac{1}{2}\sum_{i\neq j}f^{(2)}_{ij},
\end{eqnarray}
where $F^{(1)}$ and $F^{(2)}$ are the one-body and two-body operators, respectively. It is apparent that $H$ is, and should be, symmetric under the interchange $i\leftrightarrow j$~\cite{Dirac}. It is generally the case that the two-body operator $f^{(2)}_{ij}$ is symmetric under $i\leftrightarrow j$, so that we can rewrite $F^{(2)}$ as
\begin{eqnarray}
F^{(2)}=\sum_{i<j}f^{(2)}_{ij}.
\end{eqnarray}
\subsection{\textbf{The fully antisymmetric states}}
\par According to the \emph{symmetrization postulate}~\cite{Dirac}, we have to antisymmetrize the reference state to describe $N$ identical fermions. This is achieved by the standard antisymmetrization operator $\mathcal{P}_f$:
\begin{eqnarray}
 \mathcal{P}_f |\phi_{P_1}\rangle_1 |\phi_{P_2}\rangle_2 \cdots |\phi_{P_N}\rangle_N&\equiv&\sum_\pi (-1)^{\sigma(\pi)}|\phi_{P_1}\rangle_{\pi(1)}\cdots|\phi_{P_N}\rangle_{\pi(N)}\nonumber\\
 &=&\det\left(
     \begin{array}{cccc}
       |\psi_{P_1}\rangle_1 & |\psi_{P_2}\rangle_1 & \cdots & |\psi_{P_N}\rangle_1 \\
       |\psi_{P_1}\rangle_2 & |\psi_{P_2}\rangle_2 & \cdots & |\psi_{P_N}\rangle_2 \\
       \vdots & \vdots & \ddots & \vdots \\
       |\psi_{P_1}\rangle_N & |\psi_{P_2}\rangle_N & \cdots & |\psi_{P_N}\rangle_N \\
     \end{array}
   \right),
\end{eqnarray}
where the sum is over all the $N!$ permutations of the particle indices $\{1,2,\cdots,N\}$ and $\sigma(\pi)=+1$ ($-1$) if $\pi$ is an even (odd) permutation and we have written the result in the form of a determinant. Thus, the symmetrization postulate implies the Pauli exclusion principle, saying that all the $P_i$'s must be distinct if $\mathcal{P}_f |\phi_{P_1}\rangle_1 |\phi_{P_2}\rangle_2 \cdots |\phi_{P_N}\rangle_N$ describes a physical state. In turn, the SPS indices in the reference state actually satisfy $1\leq P_1<P_2<\cdots<P_N\leq M$, so that the \emph{normalized} antisymmetrized state can be written as
\begin{eqnarray}
|\psi\rangle^{(M,N)}_f\equiv\sqrt{\frac{1}{N!}}\det\left(
     \begin{array}{cccc}
       |\psi_{P_1}\rangle_1 & |\psi_{P_2}\rangle_1 & \cdots & |\psi_{P_N}\rangle_1 \\
       |\psi_{P_1}\rangle_2 & |\psi_{P_2}\rangle_2 & \cdots & |\psi_{P_N}\rangle_2 \\
       \vdots & \vdots & \ddots & \vdots \\
       |\psi_{P_1}\rangle_N & |\psi_{P_2}\rangle_N & \cdots & |\psi_{P_N}\rangle_N \\
     \end{array}
   \right).
\end{eqnarray}
Note that $|\psi\rangle^{(M,N)}_f$ is not only antisymmetric under the interchange $i\leftrightarrow j$ of two particle indices (corresponding to interchanging two rows), but also antisymmetric under the interchange $P_i\leftrightarrow P_j$ of two state indices (corresponding to interchanging two columns).
\subsection{\textbf{Fock states: filling the vacuum}}
Since the $N$ fermions are indistinguishable, we have to avoid saying ``which fermion", but to say ``\emph{how many fermions are there in a SPS}". Suppose the SPS $|\phi_\alpha\rangle$ is occupied by $N_\alpha$ (which can be either 0 or 1 and satisfy $\sum^M_{\alpha=1}N_\alpha=N$) fermions. Such a state for the $N$-fermion system can be \emph{uniquely} identified by the indices $\{P_1,\cdots,P_N\}$ of all the occupied SPSs (i.e., those with $N_\alpha=1$), and can be written as
\begin{eqnarray}
|\psi\rangle^{(M,N)}_{\mathrm{Fock}}\equiv| P_1,\cdots,P_N\rangle.
\end{eqnarray}
Note that the particles indices have disappeared in the Fock state $|\psi\rangle^{(M,N)}_{\mathrm{Fock}}$ and only the SPS indices are involved. This suggests us to introduce the following $M$ \emph{SPS creation operators}:
\begin{eqnarray}
a^\dag_\alpha,~\alpha=1,2,\cdots,M\nonumber
\end{eqnarray}
in terms of which $|\psi\rangle^{(M,N)}_{\mathrm{Fock}}$ can be generated by successively operating $a^\dag_{P_1},a^\dag_{P_2},\cdots,a^\dag_{P_N}$ onto a vacuum state $|\vec{0}\rangle\equiv|0,0,\cdots,0\rangle$ without any fermion:
\begin{eqnarray}\label{Fockaaa}
|\psi\rangle^{(M,N)}_{\mathrm{Fock}}= a^\dag_{P_1} a^\dag_{P_2} \cdots a^\dag_{P_N}|\vec{0}\rangle.
\end{eqnarray}
Note that we only used the concept of ``filling the vacuum"~\cite{Dirac} and \emph{have not yet assigned any commutation relations among the $a^\dag_\alpha$'s}.
\subsection{\textbf{Equivalence between $|\psi\rangle^{(M,N)}_f$ and $|\psi\rangle^{(M,N)}_{\mathrm{Fock}}$: relation between $a^\dag_\alpha$ and $a^\dag_\beta$}}
A key observation for the development of the formalism is the \emph{equivalence} between the fully antisymmetric state $|\psi\rangle^{(M,N)}_f$ (in the ``first-quantization language") and the Fock state $|\psi\rangle^{(M,N)}_{\mathrm{Fock}}$ (in the ``second-quantization language")~\cite{Dirac}:
\begin{eqnarray}
|\psi\rangle^{(M,N)}_f=|\psi\rangle^{(M,N)}_{\mathrm{Fock}}.
\end{eqnarray}
Since $|\psi\rangle^{(M,N)}_f$ is antisymmetric under the interchange $P_i\leftrightarrow P_j$, so must be $|\psi\rangle^{(M,N)}_{\mathrm{Fock}}$:
\begin{eqnarray}\label{Fockaaab}
|\psi\rangle^{(M,N)}_{\mathrm{Fock}}&=& a^\dag_{P_1} \cdots a^\dag_{P_i}\cdots a^\dag_{P_j} \cdots a^\dag_{P_N}|\vec{0}\rangle=-a^\dag_{P_1} \cdots a^\dag_{P_j}\cdots a^\dag_{P_i} \cdots a^\dag_{P_N}|\vec{0}\rangle.\nonumber
\end{eqnarray}
This can be achieved by imposing \emph{anticommutation relations} among the creation operators:
\begin{eqnarray}\label{adagadag}
\{a^\dag_\alpha,a^\dag_\beta\}=0,~\forall \alpha,\beta=1,2,\cdots,M,
\end{eqnarray}
and hence
\begin{eqnarray}
\{a_\alpha,a_\beta\}=0,~\forall \alpha,\beta=1,2,\cdots,M,
\end{eqnarray}
where $a_\alpha\equiv (a^\dag_\alpha)^\dag$. A direct consequence of Eq.~(\ref{adagadag}) is $(a^\dag_{\alpha})^2=0$, which is just an alternative description of the Pauli exclusion principle.
\par Having introduced the anticommutation relations among the $a^\dag_\alpha$'s, the equivalence between the antisymmetric states and the corresponding Fock state is no longer restricted to the reference state with $1\leq P_1<P_2<\cdots<P_N\leq M$. For any permutation of $(P_1,\cdots,P_N)$, say $(P'_1,\cdots,P'_N)$, we still have the following correspondence
\begin{eqnarray}\label{Pprime}
\sqrt{\frac{1}{N!}}\mathcal{P}_f |\phi_{P'_1}\rangle_1 |\phi_{P'_2}\rangle_2 \cdots |\phi_{P'_N}\rangle_N=a^\dag_{P'_1} a^\dag_{P'_2} \cdots a^\dag_{P'_N}|\vec{0}\rangle.
\end{eqnarray}
We emphasize that at this stage we only introduced the anticommutation relations among the $a^\dag_\alpha$'s and \emph{have not yet introduced any relations between $a^\dag_\alpha$ and $a_\beta$}.
\section{Derivation of the one-body operator: relation between $a^\dag_\alpha$ and $a_\beta$}
\par Our aim is to find out the corresponding one-body and two-body operators $\hat{F}^{(1)}$ and $\hat{F}^{(2)}$ in the Fock space. Due to the equivalence between $|\psi\rangle^{(M,N)}_f$ and $|\psi\rangle^{(M,N)}_{\mathrm{Fock}}$, operating $F^{(a)}$ ($a=1,2$) onto $|\psi\rangle^{(M,N)}_f$ must be equivalent to operating $\hat{F}^{(a)}$ onto $|\psi\rangle^{(M,N)}_{\mathrm{Fock}}$:
\begin{eqnarray}\label{FpsiFpsi}
F^{(a)}|\psi\rangle^{(M,N)}_f=\hat{F}^{(a)}|\psi\rangle^{(M,N)}_{\mathrm{Fock}},~a=1,2.
\end{eqnarray}
To derive the explicit form of $\hat{F}^{(a)}$ in Eq.~(\ref{FpsiFpsi}), let us calculate the left-hand-side of this equation. We first focus on the one-body operator $F^{(1)}$.
\par Note that $F^{(1)}$ ($|\psi\rangle^{(M,N)}_f$) is symmetric (antisymmetric) under the interchange of the particle labels, $i\leftrightarrow j$, the state $F^{(1)} |\psi\rangle^{(M,N)}_f$ is also antisymmetric under $i\leftrightarrow j$. Actually, it is easy to show that
\begin{eqnarray}\label{Fpsi}
F^{(1)} |\psi\rangle^{(M,N)}_f=\sqrt{\frac{1}{N!}}\sum^N_{i=1}\mathcal{P}_f[|\phi_{P_1}\rangle_1\cdots(f^{(1)}_i|\phi_{P_i}\rangle_i)\cdots|\phi_{P_N}\rangle_N],
\end{eqnarray}
which is nothing but an antisymmetrization of the new state $|\phi_ {P_1}\rangle_1\cdots(f^{(1)}_i|\phi_{P_i}\rangle_i)\cdots|\phi_{P_N}\rangle_N$, in which $|\phi_{P_i}\rangle_i$ is replaced by the new SPS $f^{(1)}_i|\phi_{P_i}\rangle_i$.
\par As an illustration, we verify Eq.~(\ref{Fpsi}) for a simple example with $M=3$ and $N=2$. In this case $|\psi\rangle^{(3,2)}_f=\sqrt{\frac{1}{2}}(|\phi_{P_1}\rangle_1|\phi_{P_2}\rangle_2-|\phi_{P_2}\rangle_1|\phi_{P_1}\rangle_2)$, so
\begin{eqnarray}
 F^{(1)} |\psi\rangle^{(3,2)}_f&=&(f^{(1)}_1+f^{(1)}_2)\sqrt{\frac{1}{2}}(|\phi_{P_1}\rangle_1|\phi_{P_2}\rangle_2-|\phi_{P_2}\rangle_1|\phi_{P_1}\rangle_2)\nonumber\\
&=&\sqrt{\frac{1}{2}}[(f^{(1)}_1|\phi_{P_1}\rangle_1)|\phi_{P_2}\rangle_2-(f^{(1)}_1|\phi_{P_2}\rangle_1)|\phi_{P_1}\rangle_2 +|\phi_{P_1}\rangle_1(f^{(1)}_2|\phi_{P_2}\rangle_2)-|\phi_{P_2}\rangle_1(f^{(1)}_2|\phi_{P_1}\rangle_2)].\nonumber
\end{eqnarray}
Note that
\begin{eqnarray}
 \mathcal{P}_f (f^{(1)}_1|\phi_{P_1}\rangle_1)|\phi_{P_2}\rangle_2&=&(f^{(1)}_1|\phi_{P_1}\rangle_1)|\phi_{P_2}\rangle_2-|\phi_{P_2}\rangle_1(f^{(1)}_2|\phi_{P_1}\rangle_2),\nonumber\\
 \mathcal{P}_f |\phi_{P_1}\rangle_1(f^{(1)}_2|\phi_{P_2}\rangle_2)&=&|\phi_{P_1}\rangle_1(f^{(1)}_2|\phi_{P_2}\rangle_2)-(f^{(1)}_1|\phi_{P_2}\rangle_1)|\phi_{P_1}\rangle_2,\nonumber
\end{eqnarray}
we thus obtain
\begin{eqnarray}
 F^{(1)} |\psi\rangle^{(3,2)}_f=\sqrt{\frac{1}{2}}[\mathcal{P}_f (f^{(1)}_1|\phi_{P_1}\rangle_1)|\phi_{P_2}\rangle_2+\mathcal{P}_f |\phi_{P_1}\rangle_1(f^{(1)}_2|\phi_{P_2}\rangle_2)],\nonumber
\end{eqnarray}
consistent with Eq.~(\ref{Fpsi}).
\par From Eq.~(\ref{Fpsi}) we see that the one-body operator $F^{(1)}$ can only change the state of a single particle. Thus, the matrix elements of $\hat{F}^{(1)}$ do not vanish only for transitions from the Fock state $|P_1,\cdots,P_i,\cdots,P_N\rangle$ to $|P_1,\cdots,P'_i,\cdots,P_N\rangle$, say. As pointed out by Landau and Lifshitz~\cite{Landau}, ``\emph{The calculation of these matrix elements is in principle very simple; it is easier to do it oneself than to follow an account of it}". Nevertheless, we believe this statement might be obscure for a beginner.
\par To proceed, we expand the SPS $f^{(1)}_i|\phi_{P_i}\rangle_i$ in Eq.~(\ref{Fpsi}) in terms of the complete set of the SPSs for particle $i$:
\begin{eqnarray}\label{Fpsi1}
F^{(1)} |\psi\rangle^{(M,N)}_f=\sqrt{\frac{1}{N!}}\sum^N_{i=1}\mathcal{P}_f\left[|\phi_{P_1}\rangle_1\cdots\left(\sum^M_{Q_i=1}~_i\langle\phi_{Q_i}|f^{(1)}_i|\phi_{P_i}\rangle_i|\phi_{Q_i}\rangle_i\right)\cdots|\phi_{P_N}\rangle_N\right].
\end{eqnarray}
It is now important observe that the matrix element $~_i\langle\phi_{Q_i}|f^{(1)}_i|\phi_{P_i}\rangle_i$ actually \emph{depends only on the state indices} $Q_i$ and $P_i$, but not on the particle label $i$. To see this, we insert the completeness relation $\int d^3x|\vec{x}_i\rangle\langle\vec{x}_i|=1_i$ for particle $i$ and get
\begin{eqnarray}
~_i\langle\phi_{Q_i}|f^{(1)}_i|\phi_{P_i}\rangle_i&=&\int\int d^3x_i d^3x'_i~_i\langle\phi_{Q_i}|\vec{x}_i\rangle\langle \vec{x}_i|f^{(1)}_i|\vec{x}'_i\rangle\langle\vec{x}'_i|\phi_{P_i}\rangle_i\nonumber\\
&=&\int\int d^3x_i d^3x'_i  \phi^*_{Q_i}(\vec{x}_i)\langle \vec{x}_i|f^{(1)}_i|\vec{x}'_i\rangle \phi_{P_i}(\vec{x}'_i)\nonumber\\
&=&\int\int d^3x  d^3x'  \phi^*_{Q_i}(\vec{x} )\langle \vec{x} |f^{(1)} |\vec{x}' \rangle \phi_{P_i}(\vec{x}' ),
\end{eqnarray}
where in the last line we have dropped out the particle index $i$ from $\vec{x}_i,~\vec{x}'_i$, and $f^{(1)}_i$. It is clear that $~_i\langle\phi_{Q_i}|f^{(1)}_i|\phi_{P_i}\rangle_i$ is the \emph{same} for all the particles and we can write
\begin{eqnarray}\label{phifphi}
~_i\langle\phi_{Q_i}|f^{(1)}_i|\phi_{P_i}\rangle_i=\langle\phi_{Q_i}|f^{(1)}|\phi_{P_i}\rangle.
\end{eqnarray}
By inserting Eq.~(\ref{phifphi}) into Eq.~(\ref{Fpsi1}) and note that the matrix element $\langle\phi_{Q_i}|f^{(1)}|\phi_{P_i}\rangle$, which has noting to do with the particle indices, is \emph{not affected} by the antisymmetrization operator $\mathcal{P}_f$, we obtain
\begin{eqnarray}\label{Fpsi2}
F^{(1)} |\psi\rangle^{(M,N)}_f
&=&\sum^N_{i=1}\sum^M_{Q_i=1}\langle\phi_{Q_i}|f^{(1)}|\phi_{P_i}\rangle\sqrt{\frac{1}{N!}}\mathcal{P}_f\left[|\phi_{P_1}\rangle_1\cdots\left(|\phi_{Q_i}\rangle_i\right)\cdots|\phi_{P_N}\rangle_N\right]\nonumber\\
&=&\sum^N_{i=1}\sum^M_{Q_i=1}\langle\phi_{Q_i}|f^{(1)}|\phi_{P_i}\rangle a^\dag_{P_1}\cdots a^\dag_{P_{i-1}}a^\dag_{Q_i}a^\dag_{P_{i+1}}\cdots a^\dag_{P_N}|\vec{0}\rangle,
\end{eqnarray}
where we have used Eq.~(\ref{Pprime}). According to Eq.~(\ref{FpsiFpsi}), we expect the last line of Eq.~(\ref{Fpsi2}) to be equal to $\hat{F}^{(1)}|\psi\rangle^{(M,N)}_{\mathrm{Fock}}=\hat{F}^{(1)}a^\dag_{P_1} a^\dag_{P_2} \cdots a^\dag_{P_N}|\vec{0}\rangle$. It is therefore desirable to replace the $a^\dag_{Q_i}$ in Eq.~(\ref{Fpsi2}) by $a^\dag_{P_i}$ through appropriate manipulations. To this end, we have to introduce the following properties for the fermion annihilation operators:
\begin{eqnarray}
a_\alpha|\vec{0}\rangle&=&0,~\forall\alpha,\label{avac}\\
\{a_\alpha,a^\dag_\beta\}&=&\delta_{\alpha\beta},~\forall \alpha,\beta.\label{aadag}
\end{eqnarray}
Using the above properties, we can show that
\begin{eqnarray}\label{QtoP}
 a^\dag_{P_1}\cdots a^\dag_{P_{i-1}}a^\dag_{Q_i}a^\dag_{P_{i+1}}\cdots a^\dag_{P_N}|\vec{0}\rangle=a^\dag_{Q_i}a_{P_i}a^\dag_{P_1}\cdots a^\dag_{P_{i-1}}a^\dag_{P_i}a^\dag_{P_{i+1}}\cdots a^\dag_{P_N}|\vec{0}\rangle.
\end{eqnarray}
It is not difficult to prove  Eq.~(\ref{QtoP}) by using Eqs.~(\ref{adagadag}), (\ref{avac}), and (\ref{aadag}) (see \ref{AppA}). Applying Eq.~(\ref{QtoP}) in Eq.~(\ref{Fpsi2}) gives
\begin{eqnarray}\label{Fpsi3}
F^{(1)} |\psi\rangle^{(M,N)}_f&=& \sum^N_{i=1}\sum^M_{Q_i=1}\langle\phi_{Q_i}|f^{(1)}|\phi_{P_i}\rangle a^\dag_{Q_i}a_{P_i}|\psi\rangle^{(M,N)}_{\mathrm{Fock}}\nonumber\\
&=& \sum^N_{i=1}\sum^M_{\alpha=1}\langle\phi_{\alpha}|f^{(1)}|\phi_{P_i}\rangle a^\dag_{\alpha}a_{P_i}|\psi\rangle^{(M,N)}_{\mathrm{Fock}},
\end{eqnarray}
where we have changed the summation index $Q_i$ into $\alpha$ as it runs over all the SPSs.
Note now that the sum over $i$ in the last line of Eq.~(\ref{Fpsi3}) can also be extended to \emph{all} the SPSs since $a_{P'_i}|\psi\rangle^{(M,N)}_{\mathrm{Fock}}=0$ if $P'_i\notin\{P_1,\cdots,P_N\}$. So,
\begin{eqnarray}\label{Fpsi4}
F^{(1)} |\psi\rangle^{(M,N)}_f&=&\sum^M_{\beta=1}\sum^M_{\alpha=1}\langle\phi_{\alpha}|f^{(1)}|\phi_{\beta}\rangle a^\dag_{\alpha}a_{\beta}|\psi\rangle^{(M,N)}_{\mathrm{Fock}}. 
\end{eqnarray}
By comparing the above equation with Eq.~(\ref{FpsiFpsi}), we finally obtain the explicit form of the one-body operator $\hat{F}^{(1)}$:
\begin{eqnarray}\label{F1}
\hat{F}^{(1)}&=&\sum^M_{\alpha,\beta=1}\langle\phi_{\alpha}|f^{(1)}|\phi_{\beta}\rangle a^\dag_{\alpha}a_{\beta}.
\end{eqnarray}
This completes our derivation of the one-body operator with the help of the newly introduced relations given by Eqs.~(\ref{avac}) and (\ref{aadag}).
\section{Derivation of the two-body operator: a straightforward extension}
\par The derivation of the two-body operator $\hat{F}^{(2)}$ closely follows that of the one-body operator. Similar to the action of $F^{(1)}$ onto $|\psi\rangle^{(M,N)}_f$, the state $F^{(2)}|\psi\rangle^{(M,N)}_f$ is antisymmetric under the interchange of the particle indices and has the form
\begin{eqnarray}\label{F2psi}
&&F^{(2)} |\psi\rangle^{(M,N)}_f=\sqrt{\frac{1}{N!}}\sum_{i<l}\mathcal{P}_f f^{(2)}_{il}[|\phi_{P_1}\rangle_1\cdots(|\phi_{P_i}\rangle_i)\cdots(|\phi_{P_l}\rangle_l)\cdots|\phi_{P_N}\rangle_N]\nonumber\\
&=&\sqrt{\frac{1}{N!}}\sum_{i<l}\mathcal{P}_f \sum^M_{Q_i=1}\sum^M_{Q_l=1}~_i\langle\phi_{Q_i}|~_l\langle\phi_{Q_l}| f^{(2)}_{il}|\phi_{P_i}\rangle_i|\phi_{P_l}\rangle_l[|\phi_{P_1}\rangle_1\cdots(|\phi_{Q_i}\rangle_i)\cdots(|\phi_{Q_l}\rangle_l)\cdots|\phi_{P_N}\rangle_N].\nonumber\\
\end{eqnarray}
Again, the matrix element $~_i\langle\phi_{Q_i}|~_l\langle\phi_{Q_l}| f^{(2)}_{il}|\phi_{P_i}\rangle_i|\phi_{P_l}\rangle_l= ~_l\langle\phi_{Q_l}|~_i\langle\phi_{Q_i}| f^{(2)}_{li}|\phi_{P_l}\rangle_l|\phi_{P_i}\rangle_i$
depends only on the state indices and can be written as
\begin{eqnarray}\label{ME21}
&&\langle\phi_{Q_i}|\langle\phi_{Q_l}| f^{(2)}|\phi_{P_i}\rangle|\phi_{P_l}\rangle=\langle\phi_{Q_l}|\langle\phi_{Q_i}| f^{(2)}|\phi_{P_l}\rangle|\phi_{P_i}\rangle.
\end{eqnarray}
Equation (\ref{F2psi}) thus becomes
\begin{eqnarray}\label{F23psi}
F^{(2)} |\psi\rangle^{(M,N)}_f&=& \sqrt{\frac{1}{N!}}\sum_{i<l}\mathcal{P}_f \sum^M_{Q_i=1}\sum^M_{Q_l=1} \langle\phi_{Q_i}| \langle\phi_{Q_l}| f^{(2)}|\phi_{P_i}\rangle |\phi_{P_l}\rangle[|\phi_{P_1}\rangle \cdots(|\phi_{Q_i}\rangle_i)\cdots(|\phi_{Q_l}\rangle_l)\cdots|\phi_{P_N}\rangle_N]\nonumber\\
&=&\sum_{i<l} \sum^M_{Q_i=1}\sum^M_{Q_l=1} \langle\phi_{Q_i}| \langle\phi_{Q_l}| f^{(2)}|\phi_{P_i}\rangle |\phi_{P_l}\rangle\sqrt{\frac{1}{N!}}\mathcal{P}_f [|\phi_{P_1}\rangle \cdots(|\phi_{Q_i}\rangle_i)\cdots(|\phi_{Q_l}\rangle_l)\cdots|\phi_{P_N}\rangle_N]\nonumber\\
&=&\sum_{i<l} \sum^M_{Q_i=1}\sum^M_{Q_l=1} \langle\phi_{Q_i}| \langle\phi_{Q_l}| f^{(2)}|\phi_{P_i}\rangle |\phi_{P_l}\rangle a^\dag_{P_1}\cdots a^\dag_{Q_i}\cdots a^\dag_{Q_l}\cdots a^\dag_{P_N}|\vec{0}\rangle.
\end{eqnarray}
Using the properties of the fermion creation and annihilation operators, we can show that (see \ref{AppB}):
\begin{eqnarray}\label{QQtoPP}
&&a^\dag_{P_1}\cdots a^\dag_{Q_i}\cdots a^\dag_{Q_l}\cdots a^\dag_{P_N}|\vec{0}\rangle=a^\dag_{Q_i}a^\dag_{Q_l}a_{P_l}a_{P_i} a^\dag_{P_1}\cdots a^\dag_{P_i}\cdots a^\dag_{P_l}\cdots a^\dag_{P_N}|\vec{0}\rangle.
\end{eqnarray}
So,
\begin{eqnarray}\label{F24psi}
F^{(2)} |\psi\rangle^{(M,N)}_f&=& \sum_{i<l} \sum^M_{Q_i=1}\sum^M_{Q_l=1} \langle\phi_{Q_i}| \langle\phi_{Q_l}| f^{(2)}|\phi_{P_i}\rangle |\phi_{P_l}\rangle a^\dag_{Q_i}a^\dag_{Q_l}a_{P_l}a_{P_i} a^\dag_{P_1}\cdots a^\dag_{P_i}\cdots a^\dag_{P_l}\cdots a^\dag_{P_N}|\vec{0}\rangle\nonumber\\
&=& \sum_{i<l} \sum^M_{Q_i=1}\sum^M_{Q_l=1} \langle\phi_{Q_i}| \langle\phi_{Q_l}| f^{(2)}|\phi_{P_i}\rangle |\phi_{P_l}\rangle a^\dag_{Q_i}a^\dag_{Q_l}a_{P_l}a_{P_i} |\psi\rangle^{(M,N)}_{\mathrm{Fock}}.
\end{eqnarray}
As before, we let $Q_i\to\alpha$, $Q_l\to\beta$ and extend the summation over $i$ and $l$ to all the SPSs to get
\begin{eqnarray}\label{F25psi}
&&F^{(2)} |\psi\rangle^{(M,N)}_f=\sum_{\gamma<\delta} \sum^M_{\alpha=1}\sum^M_{\beta=1} \langle\phi_{\alpha}| \langle\phi_{\beta}| f^{(2)}|\phi_{\gamma}\rangle |\phi_{\delta}\rangle a^\dag_{\alpha}a^\dag_{\beta}a_{\delta}a_{\gamma} |\psi\rangle^{(M,N)}_{\mathrm{Fock}}\nonumber\\
&=& \frac{1}{2} \sum^M_{\alpha=1}\sum^M_{\beta=1}\left[ \sum_{\gamma<\delta} \langle\phi_{\alpha}| \langle\phi_{\beta}| f^{(2)}|\phi_{\gamma}\rangle |\phi_{\delta}\rangle a^\dag_{\alpha}a^\dag_{\beta}a_{\delta}a_{\gamma} +\sum_{\delta<\gamma} \langle\phi_{\beta}| \langle\phi_{\alpha}| f^{(2)}|\phi_{\delta}\rangle |\phi_{\gamma}\rangle a^\dag_{\beta}a^\dag_{\alpha}a_{\gamma}a_{\delta}\right]|\psi\rangle^{(M,N)}_{\mathrm{Fock}}\nonumber\\
&=&\frac{1}{2} \sum^M_{\alpha=1}\sum^M_{\beta=1}\left[ \sum_{\gamma<\delta} \langle\phi_{\alpha}| \langle\phi_{\beta}| f^{(2)}|\phi_{\gamma}\rangle |\phi_{\delta}\rangle a^\dag_{\alpha}a^\dag_{\beta}a_{\delta}a_{\gamma} +\sum_{\delta<\gamma} \langle\phi_{\alpha}| \langle\phi_{\beta}|f^{(2)}|\phi_{\gamma}\rangle |\phi_{\delta}\rangle a^\dag_{\alpha}a^\dag_{\beta}a_{\delta}a_{\gamma}\right]|\psi\rangle^{(M,N)}_{\mathrm{Fock}},\nonumber\\
\end{eqnarray}
where in the last line we have used Eq.~(\ref{ME21}). By noting that the term with $\delta=\gamma$ does not contribute, we finally obtain
\begin{eqnarray}\label{F26psi}
\hat{F}^{(2)} &=& \frac{1}{2}\sum^M_{\alpha\beta\gamma\delta=1}\langle\phi_{\alpha}| \langle\phi_{\beta}| f^{(2)}|\phi_{\gamma}\rangle |\phi_{\delta}\rangle a^\dag_{\alpha}a^\dag_{\beta}a_{\delta}a_{\gamma}.
\end{eqnarray}
\section{Conclusions}\label{SecV}
\label{sec-final}
\par In this note, we have provide a intuitive and less ad hoc derivation of the second-quantization formalism from the its first-quantization version. The derivation avoids invoking any priori knowledge of the algebraic solution of the simple harmonic oscillator and relies only on the concept of filling the vacuum in the Fock space, as well as the equivalence between the fully symmetrized/antisymmetrized many-particle state and the corresponding Fock space. Taking the case of Fermi statistics as an example, the anticommutation relations among the fermionic creation operators naturally arise from the antisymmetrization property of the many-fermion state in first quantization. The commutation relations among the creation and annihilation operators are imposed by demanding that operating the one-body operator $F^{(1)}$ onto the antisymmetric state $|\psi\rangle^{(M,N)}_f$ recovers the action of $\hat{F}^{(1)}$ onto the Fock state $|\psi\rangle^{(M,N)}_{\mathrm{Fock}}$. The derivation of the two-body operator turns out to be a straightforward extension of the foregoing constructions.
\par We believe the method presented in this note would offer a clear and transparent interpretation of the second quantization formalism that would be friendly to graduate students or beginners who first study this topic.

  \appendix
\section{Proof of Eq.~(\ref{QtoP})}\label{AppA}
We start with the left-hand-side of Eq.~(\ref{QtoP}):
\begin{eqnarray}
&&a^\dag_{P_1}\cdots a^\dag_{P_{i-1}}a^\dag_{Q_i}a^\dag_{P_{i+1}}\cdots a^\dag_{P_N}|\vec{0}\rangle\nonumber\\
&=&(-1)^{i-1}a^\dag_{Q_i}a^\dag_{P_1}\cdots a^\dag_{P_{i-1}}a^\dag_{P_{i+1}}\cdots a^\dag_{P_N}|\vec{0}\rangle\nonumber\\
&=&(-1)^{i-1}a^\dag_{Q_i}a^\dag_{P_1}\cdots a^\dag_{P_{i-1}}(1-a^\dag_{P_{i}}a_{P_i})a^\dag_{P_{i+1}}\cdots a^\dag_{P_N}|\vec{0}\rangle\nonumber\\
&=&(-1)^{i-1}a^\dag_{Q_i}a^\dag_{P_1}\cdots a^\dag_{P_{i-1}} a_{P_i}a^\dag_{P_{i}} a^\dag_{P_{i+1}}\cdots a^\dag_{P_N}|\vec{0}\rangle\nonumber\\
&=&(-1)^{i-1}(-1)^{i-1}a^\dag_{Q_i}a_{P_i}a^\dag_{P_1}\cdots a^\dag_{P_{i-1}}a^\dag_{P_{i}} a^\dag_{P_{i+1}}\cdots a^\dag_{P_N}|\vec{0}\rangle\nonumber\\
&=&a^\dag_{Q_i}a_{P_i}a^\dag_{P_1}\cdots a^\dag_{P_{i-1}}a^\dag_{P_{i}} a^\dag_{P_{i+1}}\cdots a^\dag_{P_N}|\vec{0}\rangle.
\end{eqnarray}
\section{Proof of Eq.~(\ref{QQtoPP})}\label{AppB}
We start with the left-hand-side of Eq.~(\ref{QQtoPP}):
\begin{eqnarray}
&&a^\dag_{P_1}\cdots a^\dag_{Q_i}\cdots a^\dag_{Q_l}\cdots a^\dag_{P_N}|\vec{0}\rangle\nonumber\\
&=&(-1)^{i-1}a^\dag_{Q_i}a^\dag_{P_1}\cdots a^\dag_{P_{i-1}}a^\dag_{P_{i+1}}\cdots a^\dag_{Q_l}\cdots a^\dag_{P_N}|\vec{0}\rangle\nonumber\\
&=&(-1)^{i-1}(-1)^{l-2}a^\dag_{Q_i}a^\dag_{Q_l}a^\dag_{P_1}\cdots a^\dag_{P_{i-1}}a^\dag_{P_{i+1}}\cdots a^\dag_{P_{l-1}}a^\dag_{P_{l+1}}\cdots a^\dag_{P_N}|\vec{0}\rangle\nonumber\\
&=&(-1)^{i-1}(-1)^{l-2}a^\dag_{Q_i}a^\dag_{Q_l}a^\dag_{P_1}\cdots a^\dag_{P_{i-1}}(a_{P_i}a^\dag_{P_i}+a^\dag_{P_i}a_{P_i})a^\dag_{P_{i+1}}\cdots a^\dag_{P_{l-1}}(a_{P_l}a^\dag_{P_l}+a^\dag_{P_l}a_{P_l})a^\dag_{P_{l+1}}\cdots a^\dag_{P_N}|\vec{0}\rangle\nonumber\\
&=&(-1)^{i-1}(-1)^{l-2}a^\dag_{Q_i}a^\dag_{Q_l}a^\dag_{P_1}\cdots a^\dag_{P_{i-1}}(a_{P_i}a^\dag_{P_i})a^\dag_{P_{i+1}}\cdots a^\dag_{P_{l-1}}(a_{P_l}a^\dag_{P_l})a^\dag_{P_{l+1}}\cdots a^\dag_{P_N}|\vec{0}\rangle\nonumber\\
&=&(-1)^{i-1}(-1)^{l-2}(-1)^{i-1}a^\dag_{Q_i}a^\dag_{Q_l}a_{P_i}a^\dag_{P_1}\cdots a^\dag_{P_{i-1}}(a^\dag_{P_i})a^\dag_{P_{i+1}}\cdots a^\dag_{P_{l-1}}(a_{P_l}a^\dag_{P_l})a^\dag_{P_{l+1}}\cdots a^\dag_{P_N}|\vec{0}\rangle\nonumber\\
&=&(-1)^{i-1}(-1)^{l-2}(-1)^{i-1}(-1)^{l-1}a^\dag_{Q_i}a^\dag_{Q_l}a_{P_i}a_{P_l}a^\dag_{P_1}\cdots a^\dag_{P_{i-1}}(a^\dag_{P_i})a^\dag_{P_{i+1}}\cdots a^\dag_{P_{l-1}}(a^\dag_{P_l})a^\dag_{P_{l+1}}\cdots a^\dag_{P_N}|\vec{0}\rangle\nonumber\\
&=&-a^\dag_{Q_i}a^\dag_{Q_l}a_{P_i}a_{P_l}a^\dag_{P_1}\cdots a^\dag_{P_{i-1}}(a^\dag_{P_i})a^\dag_{P_{i+1}}\cdots a^\dag_{P_{l-1}}(a^\dag_{P_l})a^\dag_{P_{l+1}}\cdots a^\dag_{P_N}|\vec{0}\rangle\nonumber\\
&=& a^\dag_{Q_i}a^\dag_{Q_l}a_{P_l}a_{P_i}a^\dag_{P_1}\cdots a^\dag_{P_{i-1}}(a^\dag_{P_i})a^\dag_{P_{i+1}}\cdots a^\dag_{P_{l-1}}(a^\dag_{P_l})a^\dag_{P_{l+1}}\cdots a^\dag_{P_N}|\vec{0}\rangle.
\end{eqnarray}

{}


\section*{References}
\begin{thebibliography}{99}
\bibitem{Dirac} P. A. M. Dirac, \emph{Principles of Quantum Mechanics} (Oxford University Press, 1982).
\bibitem{Sakurai} J. J. Sakurai, J. Napolitano, et al., \emph{Modern quantum mechanics}, Vol. 185 (Pearson, Harlow, 2014).
\bibitem{Gottfried} K. Gottfried, T.-M. Yan, \emph{Quantum Mechanics: Fundamentals} (Springer, New York, 2003).
\bibitem{Ballentine}  L. E. Ballentine, \emph{Quantum Mechanics: A Modern Development}, Reprint 2nd ed. (World Scientific, Singapore, 2010).
\bibitem{Baym} G. Baym, \emph{Lectures on Quantum Mechanics} (Westview Press, 1974).
\bibitem{Landau} L. D. Landau and E. M. Lifshitz, \emph{Quantum mechanics: Non-relativistic theory}, 3rd ed., Vol. 3 (ButterworthHeinemann, 1981).


\end{thebibliography}
\end{document}